\documentclass[amsfonts,showpacs,tightenlines,aps,preprint,12pt,floatfix,nofootinbib]{revtex4}
\usepackage{bm}
\usepackage{epsfig}
\usepackage{float}
\usepackage{color}
\begin{document}
\preprint{ADP-17-07/T1013, YUPP-I/E-KM-17-02-2}
\thispagestyle{empty}

\title{A resolution of the inclusive flavor-breaking 
$\tau$ $\vert V_{us}\vert$ puzzle}

\author{Renwick J. Hudspith}
\email[]{renwick.james.hudspith@gmail.com}
\affiliation{Department of Physics and Astronomy, York University,
4700 Keele St., Toronto, ON CANADA M3J 1P3}
\author{Randy Lewis}
\email[]{randy.lewis@yorku.ca}
\affiliation{Department of Physics and Astronomy, York University,
4700 Keele St., Toronto, ON CANADA M3J 1P3}
\author{Kim Maltman}
\email[]{kmaltman@yorku.ca}
\affiliation{Department of Mathematics and Statistics, York University,
4700 Keele St., Toronto, ON CANADA M3J 1P3}
\altaffiliation{Alternate address: CSSM, Department of Physics,
University of Adelaide, Adelaide, SA 5005 AUSTRALIA}
\author{James Zanotti}
\email[]{james.zanotti@adelaide.edu.au}
\affiliation{CSSM, Department of Physics, University of Adelaide, 
Adelaide, SA 5005 AUSTRALIA}

\begin{abstract}
We revisit the puzzle of $\vert V_{us}\vert$ values obtained from the 
conventional implementation of hadronic-$\tau$-decay-based flavor-breaking 
finite-energy sum rules lying $>3\sigma$ below the expectations of 
three-family unitarity. Significant unphysical dependences of 
$\vert V_{us}\vert$ on the choice of weight, $w$, and upper limit, $s_0$, 
of the experimental spectral integrals entering the analysis are confirmed, 
and a breakdown of assumptions made in estimating higher dimension, $D>4$, 
OPE contributions identified as the main source of these problems. A 
combination of continuum and lattice results is shown to suggest a new 
implementation of the flavor-breaking sum rule approach in which not only 
$\vert V_{us}\vert$, but also $D>4$ effective condensates, are fit to data. 
Lattice results are also used to clarify how to reliably treat the slowly 
converging $D=2$ OPE series. The new sum rule implementation is shown to cure 
the problems of the unphysical $w$- and $s_0$-dependence of 
$\vert V_{us}\vert$ and to produce results $\sim 0.0020$ higher than those 
of the conventional implementation employing the same data.
With B-factory input, and using, in addition, dispersively constrained
results for the $K\pi$ branching fractions,
we find $\vert V_{us}\vert =0.2231(27)_{exp}(4)_{th}$, 
in excellent agreement with the result from $K_{\ell 3}$, and compatible 
within errors with the expectations of three-family unitarity, thus 
resolving the long-standing inclusive $\tau$ $\vert V_{us}\vert$ puzzle.
\end{abstract}
\pacs{12.15.Hh,11.55.Hx,12.38.Gc}

\maketitle

\section{\label{intro}Introduction}
With $\vert V_{ud}\vert = 0.97417(21)$~\cite{ht14} as input and
$\vert V_{ub}\vert$ negligible, 3-family unitary implies 
$\vert V_{us}\vert = 0.2258(9)$. Direct determinations of 
$\vert V_{us}\vert$ from $K_{\ell 3}$ and 
$\Gamma [K_{\mu 2}]/\Gamma [\pi_{\mu 2}]$, using recent 
2014 FlaviaNet experimental results~\cite{kell3andKratiosvus} and 2016 
lattice input~\cite{flag2016} for $f_+(0)$ and $f_K/f_\pi$, respectively,
yield results, $\vert V_{us}\vert = 0.2231(9)$ and $0.2253(7)$, in 
agreement with this expectation. In contrast, 
the most recent update~\cite{hflav17} of the conventional 
implementation of the finite-energy sum rule (FESR) determination employing 
flavor-breaking (FB) combinations of inclusive strange and non-strange 
hadronic $\tau$ decay data~\cite{gamizetal}, yields 
$\vert V_{us}\vert = 0.2186(21)$, $3.1\sigma$ below 
3-family-unitarity expectations. A less discrepant, but still low, 
result, $0.2207(27)$, was obtained in Ref.~\cite{aclp13} using the
same conventional implementation but somewhat higher input $K\pi$ branching 
fractions (resulting from an analysis of $K\pi$ data imposing 
additional dispersive constraints on the timelike $K\pi$ form 
factors~\cite{aclp13}). The general FB FESR framework 
whose conventional implementation produces these low $\vert V_{us}\vert$
results is outlined below.

In the Standard Model, the differential distributions, 
$dR_{V/A;ij}/ds$, for flavor $ij=ud,\, us$, vector (V) or axial-vector 
(A) current-mediated decays, with $R_{V/A;ij}$ defined by 
$R_{V/A;ij}\, \equiv\, \Gamma 
[\tau^- \rightarrow \nu_\tau \, {\rm hadrons}_{V/A;ij}\, (\gamma )]/ 
\Gamma [\tau^- \rightarrow \nu_\tau e^- {\bar \nu}_e (\gamma)]$,
are related to the spectral functions, $\rho_{V/A;ij}^{(J)}$, 
of the $J=0,1$ scalar polarizations, $\Pi^{(J)}_{V/A;ij}$, of
the corresponding current-current two-point functions, by~\cite{tsai71}
\begin{eqnarray}
{\frac{dR_{V/A;ij}}{ds}}\, &=&\, 
{\frac{12\pi^2\vert V_{ij}\vert^2 S_{EW}}
{m_\tau^2}}\, \left[ w_\tau (y_\tau ) \rho_{V/A;ij}^{(0+1)}(s)
- w_L (y_\tau )\rho_{V/A;ij}^{(0)}(s) \right]
\nonumber\\
&\equiv&\, 
{\frac{12\pi^2\vert V_{ij}\vert^2 S_{EW}}{m_\tau^2}}\, 
\left( 1-y_\tau \right)^2\, \tilde{\rho}_{V/A;ij}(s)\, ,
\label{basictaudecay}\end{eqnarray}
where $y_\tau =s/m_\tau^2$, $w_\tau (y)=(1-y)^2(1+2y)$, $w_L(y)=2y(1-y)^2$,
$S_{EW}$ is a known short-distance electroweak correction~\cite{erler}, and 
$V_{ij}$ is the flavor $ij$ CKM matrix element. The $J=0$ spectral functions, 
$\rho^{(0)}_{A;ud,us}(s)$, are dominated by the accurately known, chirally 
unsuppressed $\pi$ and $K$ pole contributions. The remaining, continuum 
contributions to $\rho^{(0)}_{V/A;ud,us}(s)$ are $\propto (m_i\mp m_j)^2$, 
and hence negligible for $ij=ud$. For $ij=us$, they are small (though not 
totally negligible) and highly constrained, through the associated $ij=us$ 
scalar and pseudoscalar sum rules, by the known value of $m_s$, making 
possible mildly model-dependent determinations in the range $s\le m_\tau^2$ 
relevant to hadronic $\tau$ decays~\cite{jop,mksps}. Subtracting the 
resulting $J=0$ contributions from the RHS of Eq.~(\ref{basictaudecay})
yields the $J=0+1$ analogue, $dR^{(0+1)}_{V/A;ij}/ds$, of $dR_{V/A;ij}/ds$,
from which the $J=0+1$ spectral function combinations 
$\rho^{(0+1)}_{V/A;ud,us}(s)$ can be determined.

The inclusive $\tau$ determination of $\vert V_{us}\vert$ employs FB 
FESRs for the spectral function combination, 
$\Delta\rho (s)\,\equiv\, \rho^{(0+1)}_{V+A;ud}(s)\, -\, 
\rho^{(0+1)}_{V+A;us}(s)$ and associated polarization difference, 
$\Delta\Pi (Q^2) \, \equiv\,  
\Pi_{V+A;ud}^{(0+1)}(Q^2)\, -\, \Pi_{V+A;us}^{(0+1)(Q^2)}$~\cite{gamizetal},
with $Q^2\, =\, -s$. Generically, for any $s_0>0$ and any choice of analytic 
weight $w(s)$,
\begin{equation}
\int_0^{s_0}w(s) \Delta\rho (s)\, ds\, =\,
-{\frac{1}{2\pi i}}\oint_{\vert s\vert =s_0}w(s) \Delta\Pi (-s)\, ds\, .
\label{basicfesr}
\end{equation}
For large enough $s_0$, the OPE is used on the RHS. 

Defining the re-weighted integrals 
\begin{equation}
R^w_{V+A;ij}(s_0)\equiv \int_0^{s_0}ds\, {\frac{w(s)}
{w_\tau (s)}}\, {\frac{dR^{(0+1)}_{V+A;ij}(s)}{ds}}\, ,
\end{equation}
and using Eq.~(\ref{basicfesr}) to replace
the FB difference
\begin{equation}
\delta R^w_{V+A}(s_0)\, \equiv\,
{\frac{R^w_{V+A;ud}(s_0)}{\vert V_{ud}\vert^2}}
\, -\, {\frac{R^w_{V+A;us}(s_0)}{\vert V_{us}\vert^2}}\, ,
\end{equation}
with its OPE representation, one finds, solving for
$\vert V_{us}\vert$~\cite{gamizetal},
\begin{equation}
\vert V_{us}\vert = \sqrt{R^w_{V+A;us}(s_0)/\left[
{\frac{R^w_{V+A;ud}(s_0)}{\vert V_{ud}\vert^2}}
\, - \delta R^{w,OPE}_{V+A}(s_0)\right]}\, .
\label{tauvussolution}\end{equation}
The result is necessarily independent of $s_0$ and $w$ so long as all
input is reliable. Assumptions employed in evaluating 
$\delta R^{w,OPE}_{V+A}(s_0)$ can thus be tested for self-consistency 
by varying $w$ and $s_0$. OPE assumptions entering the conventional 
implementation of the FB FESR approach in fact produce $\vert V_{us}\vert$ 
displaying significant $w$- and $s_0$-dependence~\cite{kmcwvus}.

The low $\vert V_{us}\vert$ results noted above are produced 
by a conventional implementation of the general FB FESR framework,
Eq.~(\ref{tauvussolution}), in which a single $s_0$ ($s_0=m_\tau^2$) 
and single weight ($w=w_\tau$), are employed~\cite{gamizetal}. 
This restriction allows the $ij=ud$ and $us$ spectral integrals to 
be determined from the inclusive $ud$ and $us$ branching fractions 
alone, but precludes carrying out $s_0$- and $w$-independence tests. 
Since $w_\tau$ has degree $3$, $\delta R^{w_\tau,OPE}_{V+A}(s_0)$ 
receives contributions up to dimension $D=8$. While $D=2$ and $4$ 
contributions, determined by $\alpha_s$ and the quark masses and 
condensates~\cite{flag2016,bckd2ope,ckd4ope,PDG,hpqcdcondratio}, are 
known, $D>4$ contributions are not. In the conventional implementation, 
$D=6$ contributions are estimated using the vacuum saturation approximation 
(VSA) (see Ref.~\cite{pp99} for the explicit expression) and 
$D=8$ contributions neglected~\cite{gamizetal,kmcwvus}. These
assumptions are potentially dangerous since the FB V+A VSA estimate involves a 
very strong double cancellation{\footnote{A factor of $\sim 3$ reduction 
occurs when the individual $ud$ and $us$ V+A sums are formed, and a further 
factor of $\sim 6$ reduction in forming the FB $ud-us$ V+A difference. 
}}, 
and the VSA is known to be badly violated, in a channel-dependent manner, 
from studies in the non-strange sector~\cite{dv7}. 

Such assumptions can, in principle, be tested by varying $s_0$. 
Writing $D>4$ contributions to $\Delta\Pi (Q^2)$ as $\sum_{D>4} C_D/Q^D$, 
with $C_D$ the effective dimension $D$ condensate, the integrated $D=2k+2$ OPE 
contribution to the RHS of Eq.~(\ref{basicfesr}), for a polynomial weight 
$w(y)=\sum_{n=0}w_n y^n$ with $y=s/s_0$, is, up to $\alpha_s$-suppressed 
logarithmic corrections, 
\begin{equation}
-\, {\frac{1}{2\pi i}}\oint_{\vert s\vert =s_0}ds\, w(y) 
\left[\Delta\Pi (Q^2)\right]_{D=2k+2}^{OPE}\, =\, 
(-1)^k\, w_k\, {\frac{C_{2k+2}}{s_0^k}}\, .
\label{higherdope}\end{equation}
Problems with the assumptions employed for $C_6$ and $C_8$ in the
conventional implementation will thus manifest themselves as an 
unphysical $s_0$-dependence in the $\vert V_{us}\vert$ results obtained 
using weights $w(y)$ with non-zero coefficients, $w_2$ and/or $w_3$, of 
$y^2$ and $y^3$.

Another potential issue for the FB FESR approach is the slow
convergence of the $D=2$ OPE series. To four loops, neglecting 
$O(m^2_{u,d}/m^2_s)$ corrections~\cite{bckd2ope}
\begin{eqnarray}
\left[\Delta\Pi (Q^2)\right]^{OPE}_{D=2}\, &&=\, 
{\frac{3}{2\pi^2}}\,
{\frac{\bar{m}_s}{Q^2}} \left[ 1 + {\frac{7}{3}} \bar{a}
+ 19.93 \bar{a}^2+ 208.75 \bar{a}^3\right]\, ,
\label{d2form}\end{eqnarray}
where $\bar{a}=\alpha_s(Q^2)/\pi$, and $\bar{m}_s=m_s(Q^2)$, $\alpha_s(Q^2)$
are the running strange mass and coupling in the $\overline{MS}$ scheme. 
With $\bar{a}(m_\tau^2)\simeq 0.1$, the ratio of $O(\bar{a}^3)$ to
$O(\bar{a}^2)$ terms is $>1$ at the spacelike point 
on $\vert s\vert =s_0$ for all kinematically accessible $s_0$. Such
slow ``convergence'' complicates choosing an appropriate 
truncation order and estimating the associated truncation uncertainty.

No apparent convergence problem exists for the $D=4$ series, which, 
to three loops, dropping numerically tiny $O(m_q^4)$ terms, is given 
by~\cite{ckd4ope}
\begin{equation}
\left[\Delta\Pi (Q^2)\right]^{OPE}_{D=4}
= {\frac{2\, \left[ \langle m_u \bar{u}u\rangle
- \langle m_s\bar{s}s\rangle\right]}{Q^4}}\, \left(
1\, -\, \bar{a}\, -\, {\frac{13}{3}}\bar{a}^2\right)\, .
\label{d4ope}\end{equation}

The slow convergence of the $D=2$ OPE series and the reliability
of conventional implementation assumptions for $C_6$ and $C_8$
will be investigated in the next section.

In the rest of the paper, the non-strange and strange spectral 
distributions entering the various FESRs considered are fixed using 
$\pi_{\mu2}$, $K_{\mu 2}$ and SM expectations 
for the $\pi$ and $K$ pole contributions, recent ALEPH data for the 
continuum $ud$ V+A distribution~\cite{aleph13}, Belle~\cite{bellekspi} 
and BaBar~\cite{babarkmpi0} results for the $K^-\pi^0$ and 
$\bar{K}^0\pi^-$ distributions, BaBar results~\cite{babarkpipiallchg} 
for the $K^-\pi^+\pi^-$ distribution, Belle results~\cite{bellekspipi} 
for the $\bar{K}^0\pi^-\pi^0$ distribution, a combination of 
BaBar~\cite{babar3K} and Belle~\cite{belle3K} results for the very 
small $\bar{K}\bar{K}K$ distribution, and 1999 ALEPH results~\cite{alephus99} 
for the combined ``residual mode'' distribution (the sum over contributions
from those strange modes not remeasured by the B-factory experiments). 
BaBar and Belle exclusive strange mode distributions are given in 
unit-normalized form, with measured branching fractions required to 
set the overall scales. We work, in general, with 2016 HFAG~\cite{HFAG2016} 
branching fractions. For the two $K\pi$ modes, however, we consider also
the alternate results, $B[\tau\rightarrow K^-\pi^0\nu_\tau ]=0.004707(181)$
and $B[\tau\rightarrow \bar{K}^0\pi^-\nu_\tau ]=0.008566(299)$, obtained in 
Ref.~\cite{aclp13} (ACLP), from an analysis imposing additional dispersive 
constraints on the timelike $K\pi$ form factors. The corresponding
2016 HFAG $K\pi$ results, obtained without the dispersive constraints, 
are $B[\tau\rightarrow K^-\pi^0\nu_\tau ]=0.004327(149)$ and 
$B[\tau\rightarrow \bar{K}^0\pi^-\nu_\tau ]=0.008386(141)$. 

\begin{center}
\begin{figure}[h]
\includegraphics[width=.60\textwidth,angle=270]
{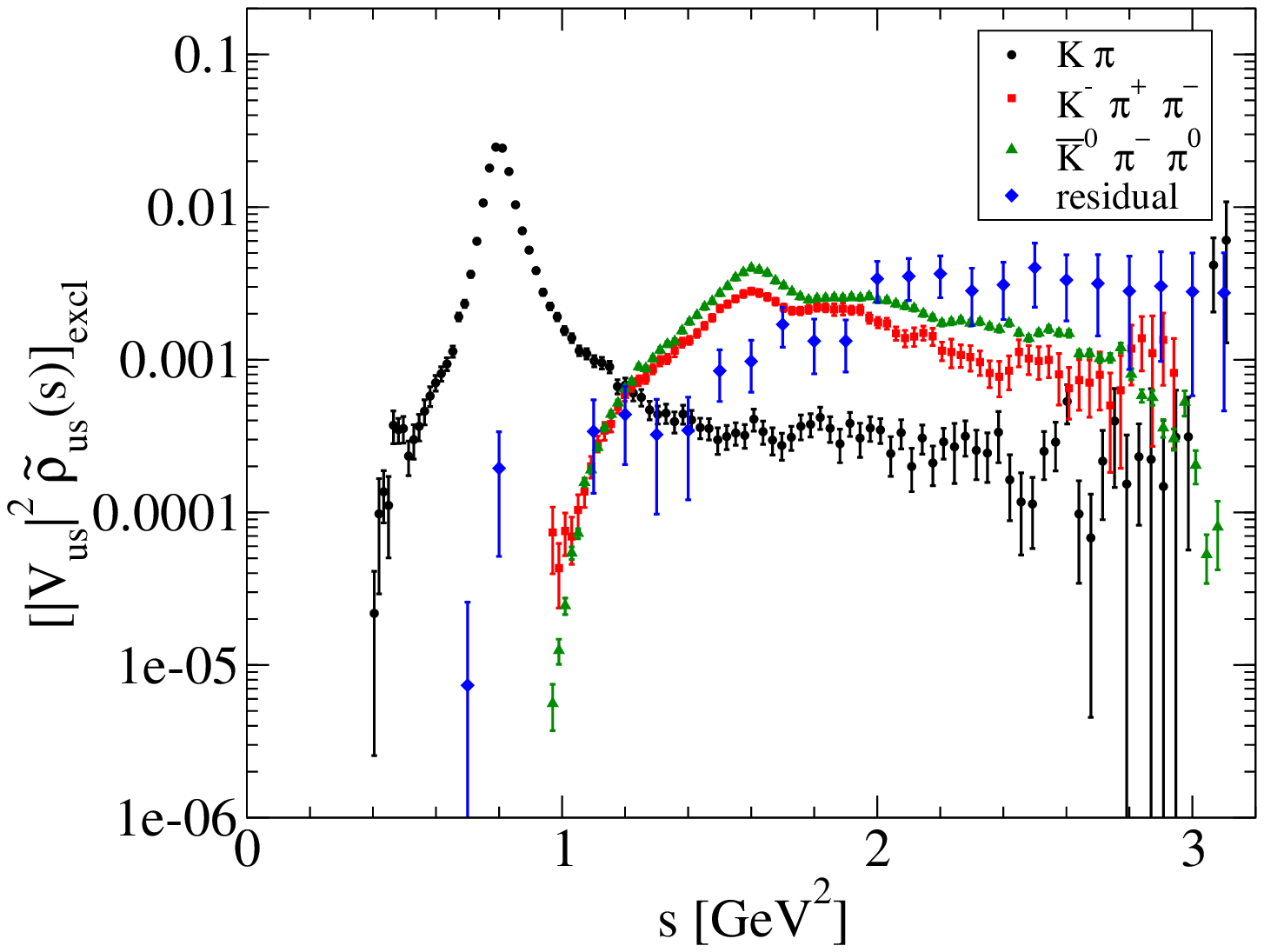}
{\caption{\label{figrhotildeus} Exclusive- and residual-mode contributions 
to the continuum $\vert V_{us}\vert^2\, \tilde{\rho}_{V+A;us}(s)$ distribution,
with 2016 HFAG normalization for the $K\pi$ points.}}
\end{figure}
\end{center}

A plot of the latest version of the ALEPH $ud$ V+A spectral distribution
may be found in Ref.~\cite{aleph13}. The exclusive- and
residual-mode contributions to the continuum $us$ V+A distribution,
in the form, $\vert V_{us}\vert^2\, \tilde{\rho}_{V+A;us}(s)$, 
directly determinable from the experiment, are shown in 
Figure~\ref{figrhotildeus}. For definiteness, the $K\pi$ points
are shown with the 2016 HFAG $K\pi$ normalization. A global rescaling
of $1.044$ is required to convert these to the alternate 2013 ACLP 
$K\pi$ normalization.

We base our central results on the additionally-constrained ACLP input choice, 
but quote results obtained using both $K\pi$ normalizations. Note that the 
publicly available ALEPH continuum $ud$ V+A distribution is normalized 
to a slightly older version of the inclusive $ud$ continuum branching 
fraction. A small rescaling ($0.5\%$ or less) is required to convert this
to the normalization implied by the branching fractions we employ.
The normalizations of the different components of the 1999 ALEPH residual 
mode distribution are also updated using HFAG 2016 branching 
fractions~\cite{HFAG2016}.  

\section{Testing conventional implementation assumptions}
The conventional implementation assumptions, $C_6\simeq C_6^{VSA}$ and
$C_8= 0$, can be efficiently investigated using appropriately
chosen $s_0$- and $w$-independence tests. A comparison of the results 
of the $w_\tau (y)=1-3y^2+2y^3$ and $\hat{w}(y)=1-3y+3y^2-y^3$ FESRs is 
particularly illuminating since the coefficients of $y^2$ in the two 
weights differ only by a sign. The corresponding integrated $D=6$ OPE 
contributions are thus identical in magnitude but opposite in sign. 
If, as the VSA estimate suggests, $D=6$ contributions are small for 
$w_\tau$, they must also be small for $\hat{w}$. Similarly, 
if integrated $D=8$ contributions are negligible for $w_\tau$, those
for $\hat{w}$, which are $-1/2$ times as large, will also be 
negligible. If conventional implementation assumptions for $C_6$ and $C_8$ 
are reliable, the $\vert V_{us}\vert$  obtained from the $w_\tau$ and 
$\hat{w}$ FESRs should thus be in good agreement, and show good individual 
$s_0$ stability. In contrast, if these assumptions are not reliable, and 
$D=6$ and $D=8$ contributions are not both small, one should see 
$s_0$-instabilities of opposite signs in the two cases, and
$s_0$-dependent differences in the results from the two FESRs which 
decrease with increasing $s_0$. 

\begin{center}
\begin{figure}[h]
\includegraphics[width=.34\textwidth,angle=270]
{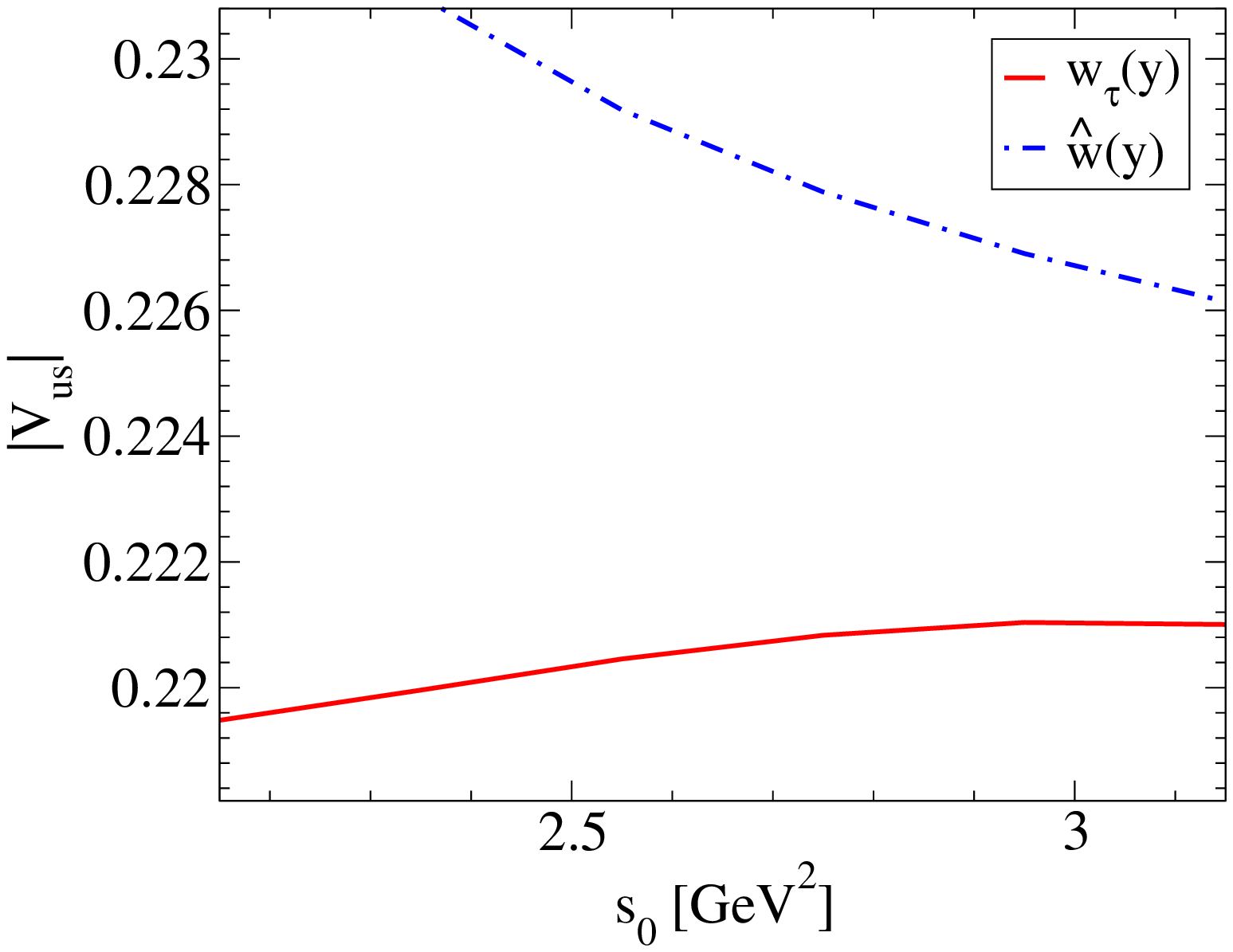}
\
\includegraphics[width=.34\textwidth,angle=270]
{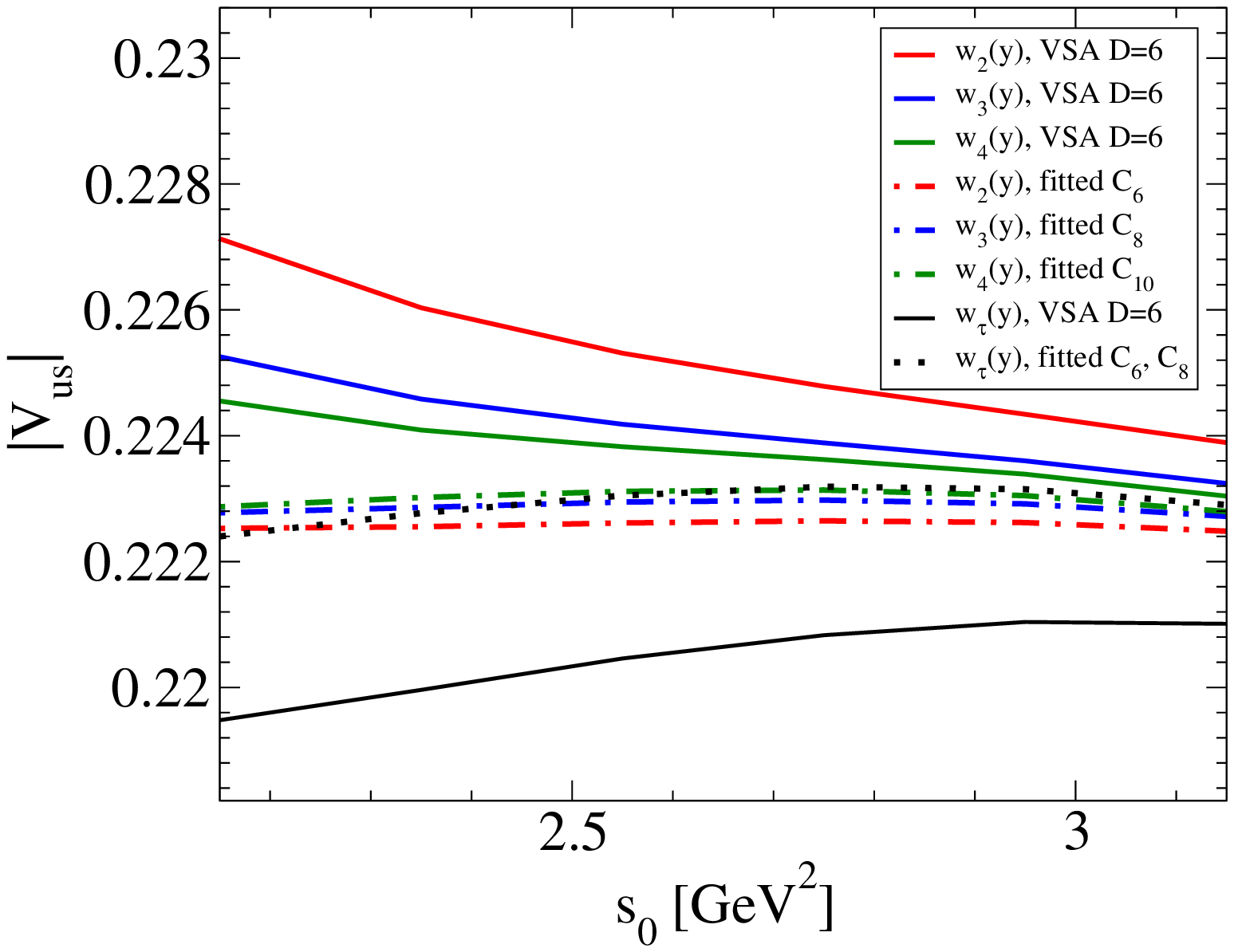}
\vskip .1in
\includegraphics[width=.34\textwidth,angle=270]
{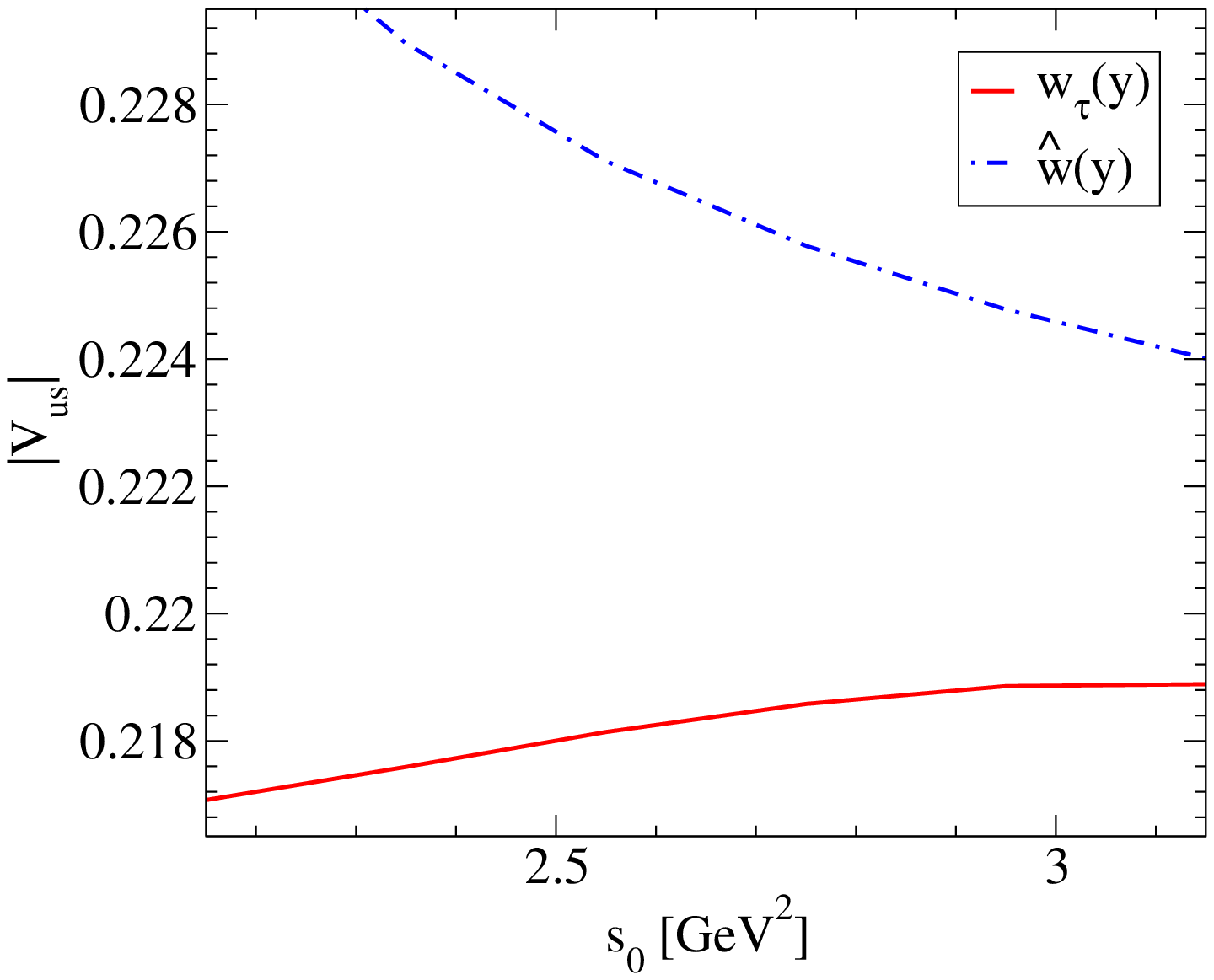}
\
\includegraphics[width=.34\textwidth,angle=270]
{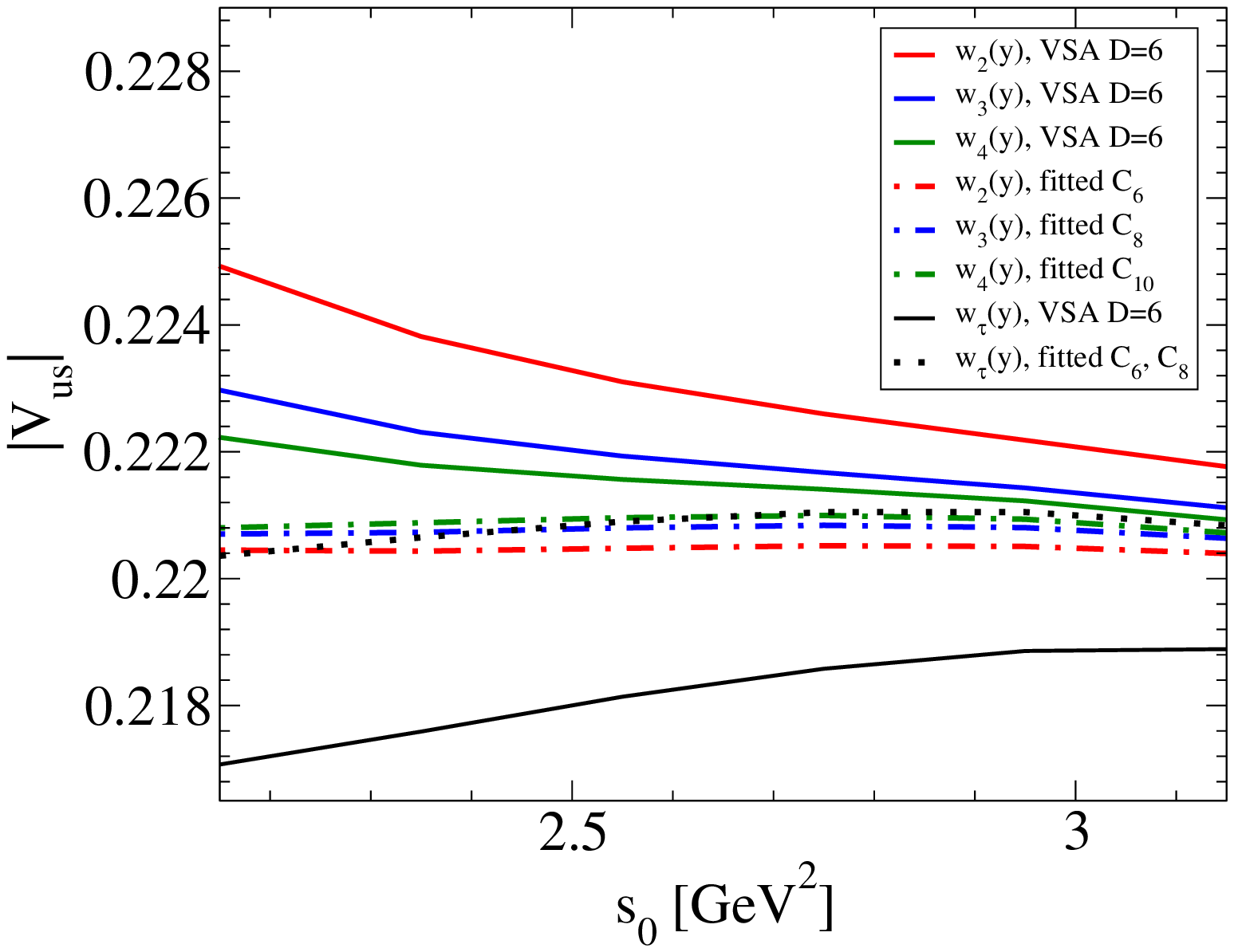}
{\caption{\label{fig1} Left panels: conventional implementation
$w_\tau$ (bottom curve) and $\hat{w}$ (top curve) results for
$\vert V_{us}\vert$. Right panels: $w_N$ and $w_\tau$ FESR results obtained 
using the fixed-order (FOPT) $D=2$ prescription. Solid lines show, top to 
bottom, conventional implementation results for $w_2$, $w_3$, $w_4$ and 
$w_\tau$. Dashed-dotted lines show, bottom to top, $w_2$, $w_3$ and $w_4$ 
results, and the dotted line $w_\tau$ results, obtained using central 
$C_{6,8}$ fit values from the alternative FB FESR analyses 
described in the text. Figures in the first row show results obtained
using the ACLP $K\pi$ normalization, those in the second row those
obtained using the HFAG $K\pi$ normalization}}
\end{figure}
\end{center}

The central values of the results of this comparison, obtained using 
the ACLP and HFAG $K\pi$ normalizations, and, to be specific, the 
3-loop-truncated contour-improved (CIPT) prescription~\cite{CIPT} for handling 
the integrated $D=2$ OPE series, are shown in the top left and bottom 
left panels of Figure~\ref{fig1}, respectively, and clearly, in both cases, 
correspond to the second scenario. One should bear in mind that the 
results for a given weight but different $s_0$ are strongly correlated, 
as are the $w_\tau$ and $\hat{w}$ results at the same $s_0$. Neither changing
the $D=2$ truncation order nor switching from CIPT to the alternate
fixed-order (FOPT) $D=2$ prescription for the $D=2$ series serves to
remove the strong, unphysical $s_0$ and weight dependences.

To understand the extent to which the $s_0$- and $w$-instabilities shown in 
Figure~\ref{fig1} are a problem for the conventional implementation $D>4$
condensate assumptions, it is useful to consider the differences between the 
$\vert V_{us}\vert$ obtained from the $\hat{w}$ and $w_\tau$ FESRs 
at the same $s_0$. If the conventional implementation assumptions are 
reliable these differences should be zero within errors. Fully propagating 
the $ud$ and $us$ experimental covariances, and adding independent 
sources of theory error in quadrature, we find, however, $\hat{w}$-$w_\tau$ 
differences of $0.0234\, (5)_{exp}\, (38)_{th}$ at $s_0=1.95\ GeV^2$,
$0.0111\, (8)_{exp}\, (21)_{th}$ at $s_0=2.55\ GeV^2$, and
$0.0064\, (16)_{exp}\, (13)_{th}$ at $s_0=3.15\ GeV^2$, clearly
signalling problems with the conventional implementation assumptions.
Similar conclusions follow from the observed $s_0$-instabilities. For
example, the difference between the $\hat{w}$ FESR results at 
$s_0=2.55\ GeV^2$ and $3.15\ GeV^2$, which should once more be zero 
within errors, is instead $0.0039\, (5)_{exp}\, (8)_{th}$. A similarly
discrepant result, $0.0096\, (8)_{exp}\, (19)_{th}$, is found for the 
difference between the $s_0=2.15\ GeV^2$ and $3.15\ GeV^2$ $\hat{w}$ results.

The top right and bottom right panels of Fig.~\ref{fig1} show the results 
of corresponding additional $w$- and $s_0$-independence tests involving the 
weights $w_N(y)$, $N=2,3,4$, with{\footnote{The $w_N(y)$, like $w_\tau (y)$, 
have a double zero at $s=s_0$ ($y=1$). This serves to keep duality violating 
contributions safely small above $s\simeq 2\ GeV^2$~\cite{pinchingdvs}.}}
\begin{equation}
w_N(y)\, =\, 1-{\frac{N}{N-1}}y+{\frac{1}{N-1}}y^N\, .
\label{wNdefn}\end{equation}
The upper solid lines in each case show the $w_2$, $w_3$ and $w_4$ results 
obtained using the conventional implementation treatment of $D>4$ OPE 
contributions and given $K\pi$ normalization, while
the dashed-dotted show lines the 
corresponding results produced by the alternate implementation discussed 
below, in which the $D>4$ effective condensates are fit to experimental data. 
The corresponding conventional and alternate implementation $w_\tau$ results 
(represented by the lowest solid and dotted lines, respectively) are 
also included for comparison. The latter are obtained using the $D=6$
and $8$ effective condensates obtained from the alternate implementation
$w_2$ and $w_3$ fits. The $s_0$-dependent, conventional implementation 
results for all of $w_\tau$, $\hat{w}$, $w_2$, $w_3$ and $w_4$ show 
evidence of converging toward a common value at $s_0>m_\tau^2$, as 
expected if the observed $s_0$-instabilities result from $D>4$ OPE 
contributions larger than those taken as input in the conventional 
implementation.

The impact of the slow convergence of the $D=2$ OPE series can be
investigated by comparing OPE expectations to lattice results for 
$\Delta\Pi (Q^2)$ over a range of Euclidean $Q^2=\, -q^2\, =\, -s$,
using variously truncated versions of the $D=2$ OPE series. Lattice 
results were obtained using the RBC/UKQCD $n_f=2+1$, 
$32^3\times 64$, $1/a=2.38$ GeV, domain wall fermion ensemble with 
$m_\pi\sim 300$ MeV~\cite{rbcukqcdfine11}. A tight cylinder cut, 
with a radius determined in a recent study of the extraction
of $\alpha_s$ from lattice current-current two-point function 
data~\cite{hlms15}, was imposed to suppress lattice artifacts at higher 
$Q^2$. The values of the light quark masses, $m_u=m_d\equiv m_\ell$ and 
$m_s$, for this ensemble, determined in Ref.~\cite{rbcukqcdfine11}, 
were used for determining the corresponding OPE expectations.

We consider the comparison first for larger $Q^2$, where $D=2$ and $4$ 
contributions should dominate. The $D=2$ OPE contribution is determined 
using ensemble values of $m_u$ and $m_s$~\cite{rbcukqcdfine11}, 
the central PDG value for $\alpha_s$~\cite{PDG}, and considering 2-, 3- 
and 4-loop truncation of the $D=2$ series. Both fixed scale, 
$\mu^2=4\ GeV^2$, and local scale, $\mu^2 =Q^2$, choices for handling 
the logarithms in the truncated series are considered. The former 
choice is the analogue of the fixed order (FOPT) prescription for 
the $D=2$ FESR contour integrals, the latter the analogue of the alternate 
CIPT prescription.{\footnote{The FOPT prescription evaluates weighted 
$D=2$ OPE integrals using a fixed-scale choice (usually $\mu^2=s_0$), 
the CIPT prescription~\cite{CIPT} using the local-scale choice,
$\mu^2=Q^2$.}} For $D=4$ contributions, Eq.~(\ref{d4ope}), we employ the
Gell-Mann-Oakes-Renner (GMOR) relation for $\langle m_u \bar{u} u\rangle$
and fix $\langle m_s\bar{s} s\rangle$ using the ensemble value
of $m_s/m_\ell$, translating the HPQCD result for 
$\langle \bar{s}s\rangle /\langle \bar{\ell}\ell\rangle$ at
physical quark masses~\cite{hpqcdcondratio}, to that for the ensemble 
masses using NLO ChPT~\cite{gl85}. 

\begin{center}
\begin{figure}[h]
\includegraphics[width=.34\textwidth,angle=270]
{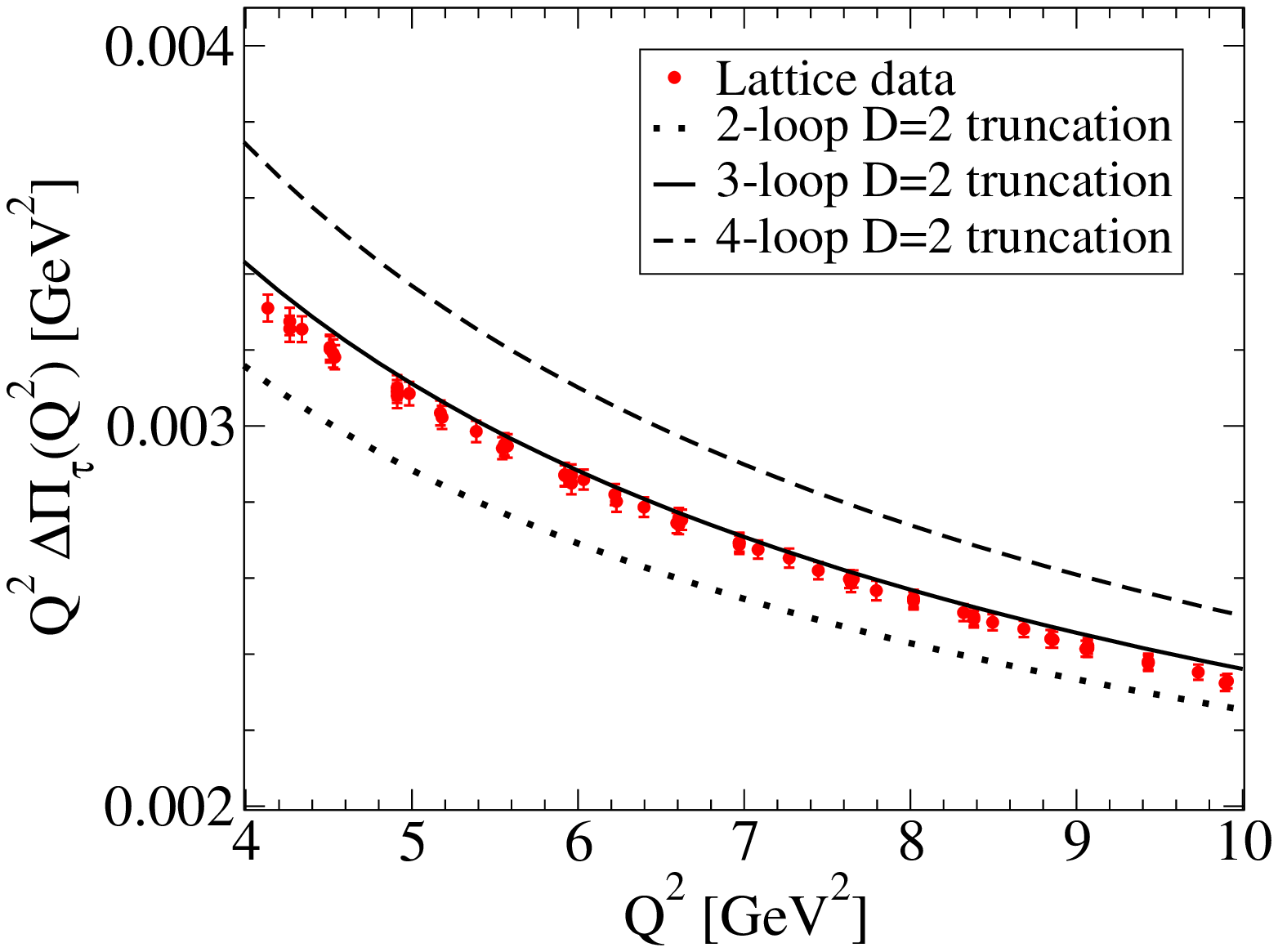}
\
\includegraphics[width=.34\textwidth,angle=270]
{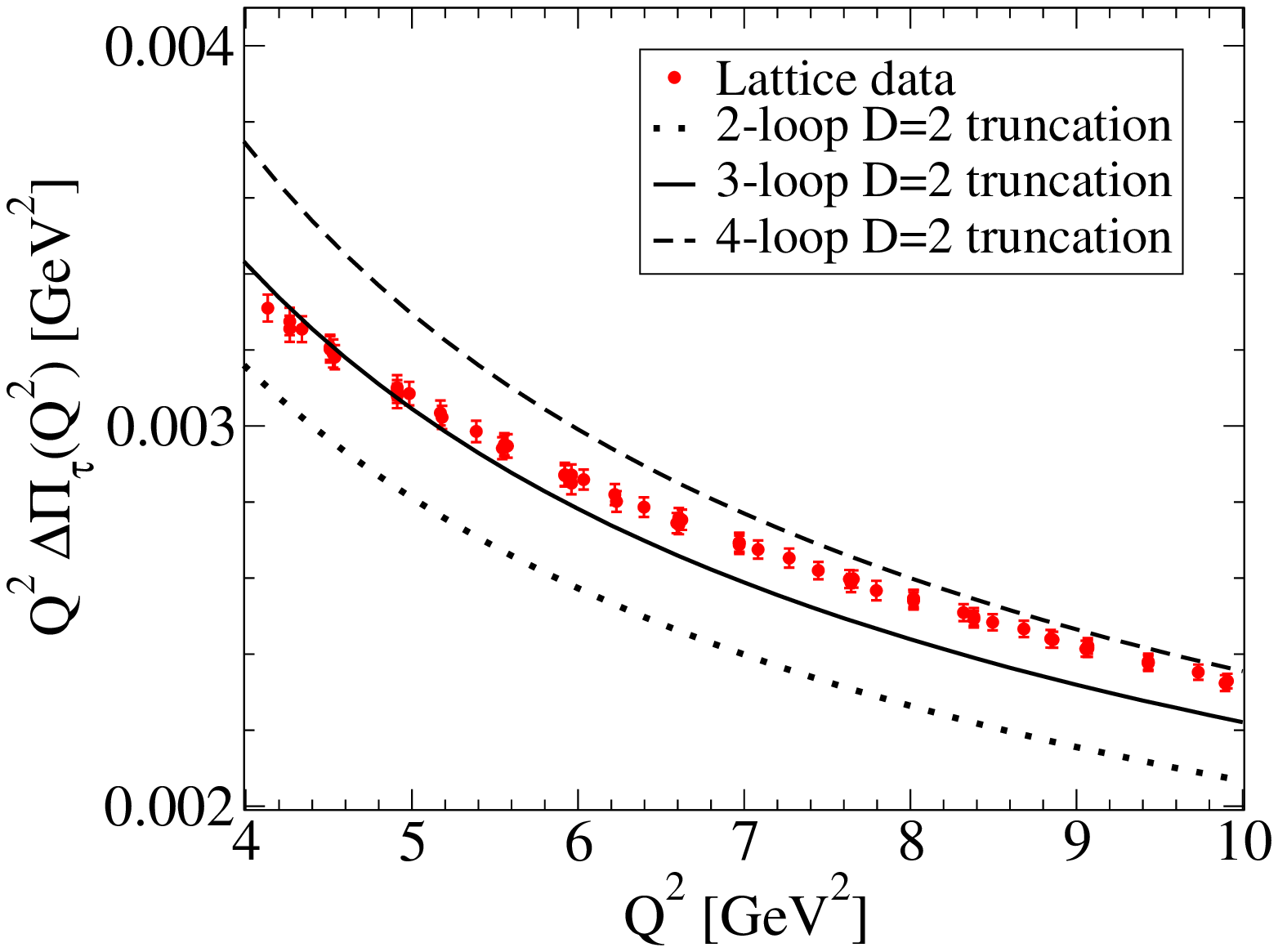}
{\caption{\label{fig2}Comparison of lattice results and $D=2+4$ OPE 
expectations for $Q^2\, \Delta \Pi_\tau (Q^2)$, for either fixed-scale 
(left panel) or local-scale (right panel) treatments of the $D=2$ series.}}
\end{figure}
\end{center}

The comparisons obtained using the fixed- and local-scale versions of the 
$D=2$ series are shown in the left and right panels of Fig.~\ref{fig2}, 
respectively. The best representation of the lattice results 
is provided by the 3-loop-truncated, fixed-scale version, which 
produces an excellent match over a wide range of $Q^2$, extending from near 
$\sim 10$ GeV$^2$ down to just above $\sim 4$ GeV$^2$, with the $Q^2$ 
dependence of the lattice results also favoring the fixed-scale over the 
alternate local-scale treatment{\footnote{Note that both the lattice 
data and OPE results at different $Q^2$ are highly correlated. These 
correlations (and not just the errors on the individual OPE and lattice 
points) must be taken into account to assess the significance (or lack 
thereof) of the difference in the $Q^2$ dependences of the local-scale 
OPE and lattice results. The uncertainty on the $Q^2$ dependence is, 
in fact, strongly dominated by that on the input strange-to-light 
condensate ratio. Taking all correlations into account, one finds,
for the fixed- and local-scale versions of the ratio of OPE to lattice 
values of the average slope between, for example, $Q^2\simeq 5\ GeV^2$ 
and $Q^2\simeq 9\ GeV^2$, the results $1.02(14)$ and $1.20(17)$, respectively, 
with $(13)$ and $(16)$ of the quoted errors coming from the uncertainty
on the input strange-to-light condensate ratio. The $Q^2$ dependence of
the lattice data thus favors the fixed-scale treatment of the $D=2$ series.
}}.

\begin{center}
\begin{figure}[h]
\includegraphics[width=.34\textwidth,angle=270]
{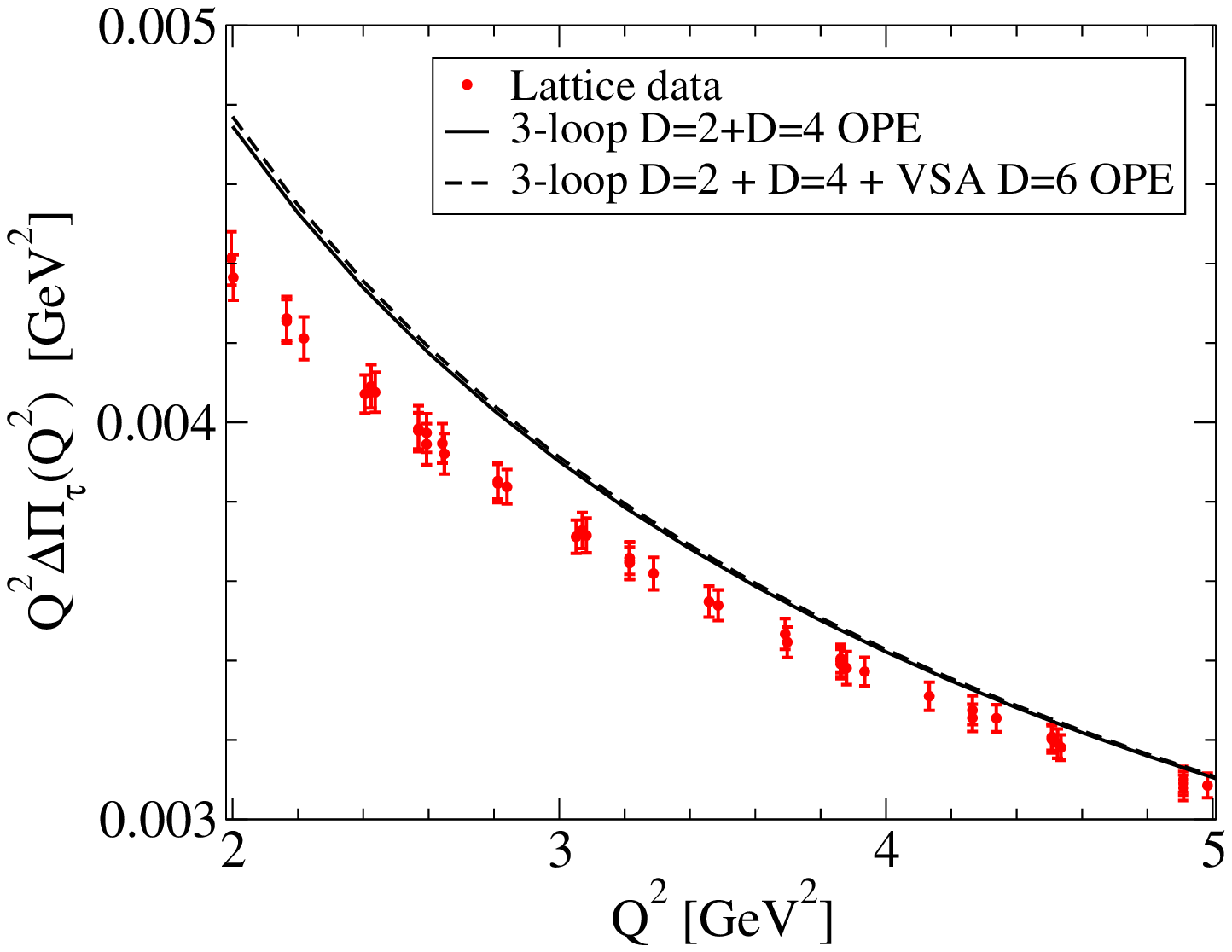}
\ 
\includegraphics[width=.34\textwidth,angle=270]
{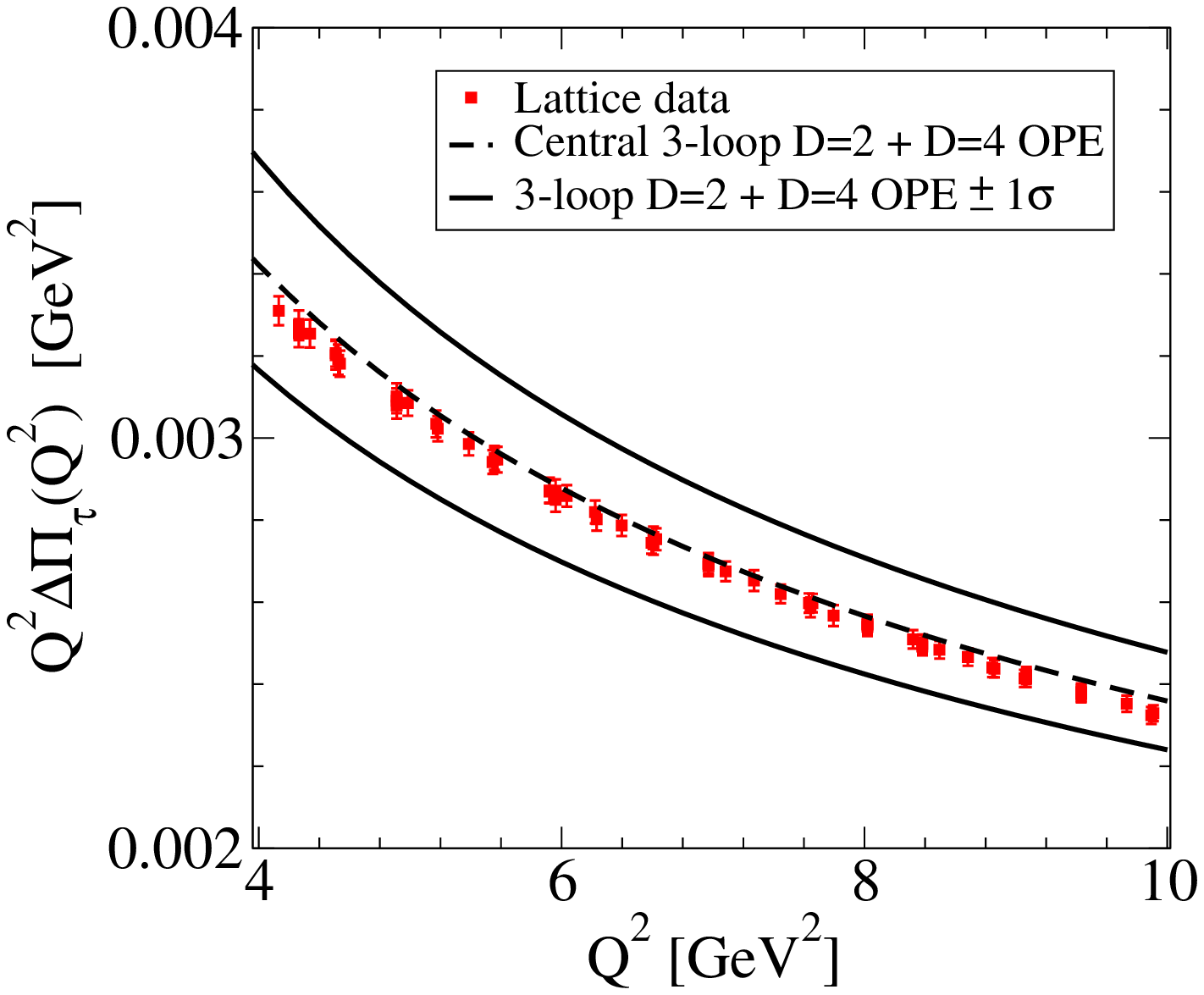}
{\caption{\label{fig3}Left panel: Comparison of lower-$Q^2$ lattice results 
to $D=2+4$ and $D=2+4+6$ OPE expectations (fixed-scale, 3-loop 
truncation for $D=2$, VSA for $D=6$).
Right panel: Lattice results and the $D=2+4$ OPE sum at larger $Q^2$,
with conventional OPE error estimates (fixed-scale, 3-loop-truncated
$D=2$).}}
\end{figure}
\end{center}

Comparison to the lattice results also provides two further useful pieces 
of information. The left panel of Fig.~\ref{fig3} shows the comparison 
of the lattice results, the three-loop-truncated, fixed-scale $D=2$ series 
version of the $D=2+4$ OPE sum, and this same $D=2+4$ OPE sum now
supplemented by the VSA estimate for $D=6$ contributions, in the
lower $Q^2$ region. Below $\sim 4\ GeV^2$, the lattice results clearly require 
$D>4$ OPE contributions much larger than those assumed in the conventional 
implementation, confirming the conclusions reached already from the 
$w_\tau$-$\hat{w}$ FESR comparison above. The right panel shows the comparison 
of the lattice results and three-loop-truncated, fixed-scale $D=2$ series 
$D=2+4$ OPE sum, now with the conventionally estimated errors for the 
latter also displayed. These are obtained by combining in quadrature 
standard estimates for the $D=4$ truncation errors with uncertainties 
produced by those on the input $D=2$ and $4$ OPE parameters. Despite
the apparently problematic convergence behavior of the $D=2$ series, 
conventional OPE error estimates are seen to provide an extremely 
conservative assessment of the uncertainty for the $D=2+4$ sum.
\section{An alternate implementation of the FB FESR approach}
The results of the previous section suggest an obvious alternative
to the conventional implementation of the FB FESR approach. First, the 
3-loop-truncated FOPT treatment favored by the comparison to the high-$Q^2$ 
lattice results is employed for the $D=2$ OPE integrals{\footnote{It is
worth noting that the prescription of truncating at 3-loop order is also 
what one would arrive at were one to interpret the series as asymptotic 
and truncate it at its smallest term.}}. Second, since both 
lattice and continuum results suggest that conventional implementation 
assumptions for the effective $D>4$ condensates, $C_D$, are unreliable, we 
avoid such assumptions and instead fit the $C_D$ to data. FESRs based on 
the weights $w_N(y)$ are particularly convenient for use in fitting the 
$C_{D>4}$ since the $w_N$-weighted OPE integral involves only a single 
$D>4$ contribution, $(-1)^N\, C_{2N+2}/\left[(N-1)\, s_0^N\right]$.
The $s_0$ dependence of the $w_N$-weighted spectral integrals in the 
region above $s_0\sim 2\ GeV^2$, where residual duality violations remain 
small, then provides sufficient information to allow both
unknowns, $\vert V_{us}\vert$ and $C_{2N+2}$, entering the $w_N$ 
FESR to be determined. 

Spectral distribution inputs employed in our analysis were outlined 
above{\footnote{Note also that, when using the ACLP $K\pi$ normalization, 
we have, for consistency, implemented the long-distance electromagnetic 
corrections employed in arriving at the $K\pi$ branching fraction results 
obtained from the ACLP analysis\cite{thanksvcep}.}}. On the OPE side, for 
the $D=2$ and $4$ contributions, we use PDG input for $\alpha_s$~\cite{PDG}, 
FLAG input for the light and strange quark masses~\cite{flag2016}, GMOR for 
the light-quark condensate~\cite{gmor}, and the HPQCD lattice 
result~\cite{hpqcdcondratio} for the ratio of strange to light quark 
condensates. The single-weight $w_2$, $w_3$ and $w_4$ FESR 
$\vert V_{us}\vert$ fit results obtained using our central (ACLP) choice 
of $K\pi$ normalization, $0.2228\, (27)_{exp}\, (4)_{th}$, 
$0.2230\, (27)_{exp}\, (4)_{th}$ and $0.2232\, (27)_{exp}\, (4)_{th}$, 
respectively, show a dramatically reduced weight dependence relative to 
those of the obtained using conventional implementation assumptions for 
the $D>4$ condensates. This is also true of the analogous results, 
$0.2205\, (23)_{exp}\, (4)_{th}$, $0.2208\, (23)_{exp}\, (4)_{th}$ and 
$0.2209\, (23)_{exp}\, (4)_{th}$, obtained using the alternate HFAG $K\pi$ 
normalization. It is worth commenting that, although the
lattice results for Euclidean $Q^2$ favor the fixed-scale treatment 
of the $D=2$ series, and hence, by extension, the FOPT prescription for the
weighted $D=2$ FESR integrals, the final results for $\vert V_{us}\vert$ 
are rather insensitive to choosing FOPT over CIPT. Explicitly, the alternate 
CIPT choice yields $0.2229\, (27)_{exp}\, (4)_{th}$ for all  of the 
$w_2$, $w_3$ and $w_4$ FESRs when the ACLP $K\pi$ normalization is used and 
$0.2206\, (23)_{exp}\, (4)_{th}$ when the HFAG $K\pi$ normalization is used. 
The CIPT treatment, of course, generates slightly different fit results 
for the $C_{D>4}$, as expected, given that FOPT and CIPT represent different 
partial resummations of the presumably asymptotic $D=2$ series.

Given the excellent consistency of the individual $w_2$, $w_3$ and $w_4$ 
FESR determinations, we take our final result from a combined 3-weight fit. 
The central ACLP $K\pi$ normalization choice yields
\begin{equation}
\vert V_{us}\vert = 0.2231\, (27)_{exp}\, (4)_{th}\, ,
\label{finalcentralvus}\end{equation}
$0.0022$ higher than the result obtained using conventional implementation 
$D>4$ assumptions with the same experimental input. This result is in 
excellent agreement with the result from $K_{\ell 3}$ and compatible 
within errors with the expectations of 3-family unitarity. The combined 
3-weight fit result, $\vert V_{us}\vert = 0.2208\, (23)_{exp}\, (4)_{th}$, 
generated by the alternate (HFAG) choice of $K\pi$ normalization, similarly, 
lies $0.0020$ above the result obtained employing the same experimental 
input and conventional implementation assumptions for the $D>4$
condensates.

Table~\ref{tab1} shows the error budgets for the $w_2$, $w_3$ and $w_4$ 
fits employing the ACLP $K\pi$ normalization. Theory errors, resulting from 
uncertainties in the input parameters $\alpha_s$, $m_s$ and 
$\langle m_s\bar{s} s\rangle$, and the small $J=0$ continuum
subtraction, are labelled by $\delta\alpha_s$,
$\delta m_s$, $\delta\langle m_s\bar{s}s\rangle$ and $\delta (J=0\ sub)$,
respectively, and shown above the horizontal line. Those induced by the 
covariances of the non-strange and strange experimental distributions 
$dR_{V+A;ud}/ds$ and $dR_{V+A;us}/ds$ are denoted $\delta^{exp}_{ud}$ 
and $\delta^{exp}_{us}$ and shown below the horizontal line. The 
$\delta_{us}^{exp}$ uncertainties strongly dominate the total 
errors.
\begin{center}
\begin{table}[h]
\caption{Single-weight fit $\vert V_{us}\vert$ error contributions for the
$w_2$, $w_3$ and $w_4$ FESRs, using 3-loop-truncated FOPT for the $D=2$ 
OPE series. Notation as described in the text.}
\vskip .05in
{\begin{tabular}{@{}lccc@{}} 
\toprule
\ \ Source&\ \ \ \ \ $\delta\vert V_{us}\vert$\ \ \ \ \  &\ \ \ \ \ $\delta\vert
V_{us}\vert$\ \ \ \ \ &\ \ \ \ \ $\delta\vert V_{us}\vert$\ \ \ \ \  \\
&$w_2$ FESR&$w_3$ FESR&$w_4$ FESR\\
\colrule
$\delta\alpha_s$&0.00002&0.00006&0.00006\\
$\delta m_s(2\ GeV)$&0.00008&0.00009&0.00008\\
$\delta\langle m_s\bar{s}s\rangle$&0.00035&0.00035&0.00035\\
$\delta (J=0\ sub)$&0.00009&0.00009&0.00009\\
\hline
$\delta_{ud}^{exp}$&0.00027&0.00028&0.00028\\
$\delta_{us}^{exp}$&0.00272&0.00273&0.00273\\
\botrule
\end{tabular}\label{tab1}}
\end{table}
\end{center}
From the lattice-OPE comparison discussed above, the estimates 
in the upper half of the table should provide a very conservative 
assessment of theoretical uncertainties. Combining the different components 
in quadrature yields a total theory error of $0.0004$ on 
$\vert V_{us}\vert$ for all three determinations. 
The new implementation of the FB FESR approach is thus competitive with 
the alternate $K_{\ell 3}$ and $\Gamma [K_{\mu 2}]/\Gamma [\pi_{\mu 2}]$ 
determinations from a theory error point of view, though improvements to 
the errors on the strange experimental distributions are required to make
it fully competitive over all.

To test whether fitting the $D>4$ condensates has solved the problem 
of the $s_0$-instabilities found in the conventional implementation, we 
have rerun the $s_0$-dependent $w_N$ analyses, using the central fitted 
$C_{2N+2}$ values as input{\footnote{It is worth noting that the
central fitted $C_{2N+2>4}$ produce contributions to the $w_N$ 
FESRs which appear natural in size relative to the known $D=2$ and
$4$ contributions. At $s_0=m_\tau^2$, for example, relative to the
corresponding $D=2$ contributions, $D=4$ and $6$ contributions are 
$\sim 83\%$ and $-26\%$ for $w_2$, $D=4$ and $8$ contributions $\sim 67\%$ 
and $-11\%$ for $w_3$, and $D=4$ and $10$ contributions $\sim 61\%$
and $-5\%$ for $w_4$.}}. 
The dashed-dotted lines in the right panel of Fig.~\ref{fig1} show the 
results of this exercise. Using the fitted $C_{D>4}$ values dramatically 
reduces the $s_0$-instabilities of the conventional implementation 
versions of the same analyses, providing a strong self-consistency 
check on the new FB FESR implementation. The dotted line in this
same panel shows the analogous $\vert V_{us}\vert$ results obtained 
from the $s_0$-dependent $w_\tau$ analysis using the fitted values 
of $C_6$ and $C_8$ as input. One again finds a dramatic reduction in 
the $s_0$ dependence, as well as excellent agreement with the results 
obtained using the other weights.

Errors on the $us$ distribution data limit the precision with which the 
$C_{2N+2}$ (which represent nuisance parameters for the determination of 
$\vert V_{us}\vert$) can be currently determined. It is, nonetheless, worth 
checking that the results for the FB condensates are compatible with an 
expected FB suppression relative to the corresponding flavor $ud$ V+A 
condensates. Comparing the results for $C_6$ and $C_8$ from our favored 
(FOPT) fits to those for the corresponding $ud$ V+A condensates, 
$C_{6,8}^{ud;V+A}$, obtained from the favored, $s_{min}=1.55\ GeV^2$, 
3-weight, combined V\&A FOPT fit of Ref.~\cite{dv7}, we find, for the 
ratios of FB to non-FB $D=6$ and $8$ condensates, the results $0.50(16)(20)$ 
and $0.40(25)(19)$, respectively, where the first error, in each case, 
results from the uncertainty on the FB condensate $C_{2N+2}$ and the 
second from that on $C_{2N+2}^{ud;V+A}$. The results for the FB condensates 
are thus natural, and compatible with the expectation of FB suppression; the 
sizeable uncertainties, however, preclude going beyond these qualitative 
observations.

\section{Conclusions}
We have revisited the determination of $\vert V_{us}\vert$ from 
flavor-breaking finite-energy sum rule analyses of experimental 
inclusive non-strange and strange hadronic $\tau$ decay distributions, 
identifying an important systematic problem in the conventional 
implementation of this approach, and developing an alternate 
implementation which cures this problem. We have also used lattice 
results to bring under better theoretical control the treatment
of the potentially problematic $D=2$ OPE series entering these analyses. 
The new implementation, which employs the FOPT prescription for the 
integrated $D=2$ OPE series and requires fitting effective $D>4$ 
condensates to data, dramatically reduces the $w$- and 
$s_0$-instabilities found when conventional implementation 
assumptions are employed for the $D=6,\, 8$ condensates. The
$w$- and $s_0$-instabilities of the conventional implementation 
establish that the assumptions employed in that implementation are 
not self-consistent, and hence that the conventional implementation 
needs to be abandoned going forward.

It is worth commenting on the relation to earlier attempts to
bring the unphysical $w$- and $s_0$-dependence of the results
for $\vert V_{us}\vert$ under improved control. Refs.~\cite{kmcw06,kmcw07} 
employed degree $8$, $10$ and $20$ weights constructed to simultaneously 
(i) emphasize $D=2$ contributions from the part of the contour with lower 
$\vert \alpha_s(Q^2)\vert$, with the goal of improving the convergence
of the $D=2$ series integrated using the CIPT prescription, and (ii) keep 
the coefficients $w_N$, $N\ge 2$ in $w(y)=\sum_N w_N y^N$, which govern 
$D>4$ OPE contributions, relatively small~\cite{kmjkaltwts}. Relative to 
the weights $w_N(y)$ employed above, the earlier weights have the 
disadvantage of producing large numbers of experimentally unconstrained $D>4$ 
contributions, several governed by coefficients larger than those 
appearing in the $w_N(y)$. Focusing on the ``ACO'' section of Table II 
of Ref.~\cite{kmcw06}, which employs strange exclusive branching fractions 
closest to (if slightly higher than) those used here, we find, not 
surprisingly, $s_0$-dependences significantly larger than those found from 
the new implementation employing the lower degree $w_N$, which choices
allow the relevant $D>4$ effective condensates to be fit to data, 
rather than neglected as in cases of the weights used in 
Refs.~\cite{kmcw06,kmcw07}. Similarly, the weight-dependence of the 
$s_0=m_\tau^2$ $\vert V_{us}\vert$ results quoted in Ref.~\cite{kmcw07} is 
significantly larger than that found from the new implementation,
quoted above. The new implementation thus also supercedes those earlier 
attempts to address the same $w$- and $s_0$-dependence problems.

The new implementation produces results for $\vert V_{us}\vert$ 
$\sim 0.0020$ higher than those obtained analyzing the same data using 
conventional implementation assumptions for the $D=6$ and $8$ condensates. 
Taking into account the additional dispersive constraints incorporated by 
the ACLP $K\pi$ normalization, we find a result, Eq.~(\ref{finalcentralvus}), 
in excellent agreement with that obtained from $K_{\ell 3}$, and compatible 
within errors with the expectations of three-family unitarity, thus 
resolving the long-standing puzzle of the $>3\sigma$ low values of 
$\vert V_{us}\vert$ obtained from the conventional implementation
of the FB FESR $\tau$ approach. 

Roughly half of the increase from the $0.2186(21)$ conventional 
implementation result for $\vert V_{us}\vert$ quoted in Ref.~\cite{hflav17} 
comes from the shift to the ACLP $K\pi$ normalization and half from the 
use of the new implementation strategy. The use of results for the
$D>4$ condensates obtained from fits to data in place of the 
non-self-consistent conventional implementation assumptions for these 
condensates is a particularly important feature of the new implementation.

The FB FESR approach to the determination of $\vert V_{us}\vert$ has been 
shown to have very favorable theory errors. The limitations,
at present, are entirely experimental in nature, with errors strongly 
dominated by those on the weighted inclusive strange spectral integrals. 
In this regard, it is worth noting that the errors on the lower-multiplicity 
{\it exclusive-mode} $K^-\pi^0$, $\bar{K}^0\pi^-$, $K^-\pi^+\pi^-$ and 
$\bar{K}^0\pi^-\pi^0$ contributions, all of which are based on the
much higher statistics BaBar and Belle distribution data are, at present, 
dominated by the uncertainties on the corresponding branching fractions 
(which normalize the unit-normalized experimental distributions). 
Significant improvements to the overall error can thus be achieved 
through improvements to the strange exclusive-mode branching 
fractions without requiring simultaneous, experimentally more difficult, 
improvements to the associated differential distributions.

\begin{acknowledgments}
Thanks to RBC/UKQCD for providing access to the data of 
Ref.~\cite{rbcukqcdfine11}, used in the OPE-lattice study of the 
conventional FB FESR implementation. Lattice propagator inversions 
were performed on the STFC-funded ``DiRAC'' BG/Q system in the 
Advanced Computing Facility at the University of Edinburgh. 
The work of R.J.H., R.L. and K.M. is supported by the Natural
Sciences and Engineering Research Council of Canada, that of J.M.Z. by 
Australian Research Council grants FT100100005 and DP140103067.
\end{acknowledgments}


\vfill\eject
\end{document}